Turkish Comments on "Future Perspectives in HEP"

Engin Arık[*]
*Department of Physics, Faculty of Arts and Sciences, Boğaziçi University, Bebek,80815,Istanbul, Turkey*

Saleh Sultansoy[**]
*Department of Physics, Faculty of Arts and Sciences, Gazi University, Teknikokullar, 06500, Ankara, Turkey*
*Institute of Physics, Academy of Sciences, H. Cavid Avenue 33, 370143, Baku, Azerbaijan*

Abstract

These comments were prepared during the ICFA Seminar on "Future Persrectives in High Energy Physics" held at CERN (8-11 October 2002) and partially presented at the Panel and General Discussion on Interregional Collaboration for Future Facilities (10 October). Comments include arguments favoring the existence of the fourth SM family and new level of compositness. Then, the possible role of the linac-ring type colliders for future HEP research is discussed emphasizing TeV scale lepton-hadron and photon-hadron colliders. This preprint presents prepared transparencies and some illuminating remarks with corresponding references.

[*] engin@ocean.phys.boun.edu.tr

[**] saleh@gazi.edu.tr

*ICFA Seminar, CERN, 10 Oct 2002*

# TURKISH COMMENTS ON
## "Future Perspectives in HEP"

Engin ARIK[1] and Saleh SULTANSOY[2,3]

[1] *Boğaziçi Univ., İstanbul, TÜRKİYE*
[2] *Gazi Univ., Ankara, TÜRKİYE*
[3] *Institute of Physics, Baku, Azerbaycan*

1. Periodic Table of Elementary Particles

2. Flavour Democracy ➔ the Fourth SM Family

3. SUSY vs Compositness ➔ SUSY at pre(pre)onic level

4. TeV scale lepton-hadron and photon-hadron colliders

5. Linac-Ring type factories: TAC Project

# 1. Periodic Table of Element(*)s
## *ary Particle

|   | $v$ | $l$ | $u$ | $d$ |
|---|---|---|---|---|
| 1 | < 3 eV | 0.510998902(21) MeV | 1.5÷4.5 MeV | 5÷8.5 MeV |
| 2 | <0.19 MeV | 105.658357(5) MeV | 1.0÷1.4 GeV | 80÷155 MeV |
| 3 | <18.2 MeV | 1776.99±0.29 MeV | 174.3±5.1 GeV | 4÷4.5 GeV |
| 4 | > 45 GeV | >100.8 GeV | >200 GeV ?? | >128 GeV (>199) |

$m_\gamma < 2 \times 10^{-16}$ eV,     $m_g = 0$ (< few MeV)

$m_W = 80.423(39)$ GeV,   $m_Z = 91.1876(21)$ GeV

$m_H > 114$ GeV

$$V^q_{CKM}(|V^q_{ij}|) = \begin{bmatrix} 0.9721-0.9747 & 0.215-0.224 & 0.02-0.05 & ... \\ 0.209-0.227 & 0.966-0.976 & 0.038-0.044 & ... \\ 0-0.09 & 0-0.12 & 0.08-0.9993 & ... \\ ... & ... & ... & ... \end{bmatrix}$$

$V^l$:    4×4  *(Dirac case)*
        8×8  *(Majorana case)*

*In our opinion the mysteries hidden in the above data present the most crucial questions to be answered in HEP*

## 2. The Fourth SM Family

*i) Flavor Democracy + 3$^{rd}$ family masses*
   *→ at least 1 extra generation should exist*

*ii) Presicion EW data*
   - *2000: the 4$^{th}$ family excluded at 99% CL*
   - *2002: 3 and 4 families have the same status*

*iii) gg→H is enhanced by a factor ~ 6÷9*
   *the "golden mode" becomes more important even at the Tevatron*

*iv) Direct production*
   - *4$^{th}$ family quarks will be copiously produced at the LHC*
   - *CLIC is the most suitable machine for the investigation of the 4$^{th}$ family leptons*

*In our opinion, 4$^{th}$ SM family will be the most striking discovery next to the Higgs boson*

# 3. SUSY at the pre(pre)onic level

*Arguments favoring the new level of compositness:*
- *Number of "fundamental" particles*
- *Family replication*
- *Interfamily (CKM) mixings – especially !!!*

*"Historical" arguments:*
- Mistery of Mendeleyev's Periodic Table of Elements was clarified after the discovery of *e, p, n*
- Mistery of "Periodic Table" of hadrons was clarified after the discovery of *quarks*
- Mistery of "Periodic Table" of elementary particles will be clarified after the discovery of *preons*

*Number of free parameters (observable !!)*

|  | 3 families |  | 4 families |  |
|---|---|---|---|---|
|  | SM | MSSM | SM | MSSM |
|  | 19 → 26 | ~160 | 40 | ~250 |

*SUSY should be realized at the most fundamental pre(pre)nic level !!*

# 4. TeV scale lepton-hadron and photon-hadron colliders

- *The only "realistic" way is linac-ring type machines*

- *They are complementary to pp and $e^+e^-$ colliders*

| Colliders | Hadron | Lepton | Lepton-Hadron |
|---|---|---|---|
| 1990's | Tevatron | SLC/LEP | HERA |
| $\sqrt{s}$, TeV | 2 | 0.1/0.1→0.2 | 0.3 |
| L, $10^{31}$ | 1 | 0.1/1 | 1 |
| | | | |
| 2010's | LHC | "NLC"(TESLA?) | "NLC"×LHC |
| $\sqrt{s}$, TeV | 14 | 0.5→1.0(0.8)  (×7) | 3.7 |
| L, $10^{31}$ | $10^3$ | $10^3$ | 1-10 |
| | | | |
| 2020's | VLHC | CLIC | CLIC×VLHC |
| $\sqrt{s}$, TeV | 200 | 3  (×11) | 34 |
| L, $10^{31}$ | $10^3$ | $10^3$ | 10-100 |

*Additional options:*   $\gamma p$, $eA$, $\gamma A$, FEL $\gamma A$

*Possible stages:*

- *e-RHIC (10 GeV e-linac)*
- *THERA (TESLA on HERA) with √s = 1÷1.6 TeV and L ~ $10^{31} cm^{-2} s^{-1}$ (see TESLA TDR)*
- *QCD Eplorer at CERN (70 GeV "CLIC" on LHC) with √s = 1.4 TeV and L ~ $10^{31} cm^{-2} s^{-1}$*

*QCD Explorer is necessary for:*

- *Adequate interpretation of the LHC data*
- *To explore the region of extremely small $x_g = 10^{-5} - 10^{-6}$ at sufficiently high $Q^2 = 1-10$ GeV$^2$*

*in addition, opportunity to test basic CLIC technologies at more relaxed conditions*

---

## 5. TAC Project

*Linac-Ring type φ and c-τ factories with L~$10^{34} cm^{-2} s^{-1}$*

*For details see:* http://bilge.science.ankara.edu.tr

# *Appendix: Flavor Democracy*

*It is useful to consider three different bases: Standard Model basis $\{f^0\}$, Mass basis $\{f^m\}$ and Weak basis $\{f^w\}$.*

$L_Y^{(d)} = \Sigma\, a_{ij}^d\, (u^0_{Li}\, d^0_{Li}) \begin{pmatrix} \varphi^+ \\ \varphi^- \end{pmatrix} d^0_{Ri} + h.c. = \Sigma\, a_{ij}^d\, d^0_i\, d^0_j\,;$

$m_{ij}^d = a_{ij}^d\, \eta$    (i,j = 1, .. , n, where n is the number of SM families)

*Diagonalization of mass matrix, $\{f^0\} \rightarrow \{f^m\}$ by bi-unitary transformation.*

*Then,*    $U_{CKM} = (U^u_L)^+ U^d_L$      *$\{f^w\}$ basis*

## *Flavor Democracy Assumptions*

*First Assumption: before SSB there are no differences between $d^0, s^0, b^0,..$
Therefore,*

$a_{ij}^d \approx a^d\,,\ \ a_{ij}^u \approx a^u\,,\ \ a_{ij}^l \approx a^l\,,\ \ a_{ij}^\nu \approx a^\nu$

*$\rightarrow$ n-1 massless particles and one massive with $m = na^F \eta$  (F=u,d,l,$\nu$)*

*Second Assumption: there is only one Higgs doublet. Therefore,*

$a^d \approx a^u \approx a^l \approx a^\nu$

*In terms of mass matrix above arguments mean:*

$$M^0 = a\eta \begin{pmatrix} 1 & 1 & 1 & 1 \\ 1 & 1 & 1 & 1 \\ 1 & 1 & 1 & 1 \\ 1 & 1 & 1 & 1 \end{pmatrix} \rightarrow M^m = 4\, a\eta \begin{pmatrix} 0 & 0 & 0 & 0 \\ 0 & 0 & 0 & 0 \\ 0 & 0 & 0 & 0 \\ 0 & 0 & 0 & 1 \end{pmatrix}$$

*Third Assumption: a is between $e = g_W \sin\theta_W$ and $g_Z = g_W/\cos\theta_W$. Therefore, the fourth SM famiky fermions are almost degenerate and $320 < m_4 < 730$ GeV.*

*Small deviations from flavor democracy $\rightarrow$ masses of first three family fermions and CKM mixings*

Below we present some illuminating remarks on topics covered by our Comments and corresponding references. It should be noted that a list of references is far from completeness and any suggestion on this subject is welcomed. We refer mainly to review articles on each topics and add a few articles followed the latest review.

1. Flavor Democracy and the Fourth SM Family

The flavor democracy hypothesis (or, in other words, democratic mass matrix approach) was introduced in seventies taking in mind three SM families. Later, this idea was disfavored by the large value of the top quark mass. In nineties the hypothesis was revisited assuming that extra SM families exist (the LEP data on neutrino counting is valid only for those with masses below half of the Z boson mass). The situation in year 2000, as well as wide list of corresponding references, can be found in [1].

The most important development since 2000 connected to presicion electroweak data: it was shown [2] that according to the latest data three and four SM family cases has similar status. The existence of the fourth family will yield an essential enhancement in Higgs boson production, via gluon-gluon fusion at Tevatron and LHC [3]. Then, due to small mixing of the fourth SM family fermions with those of the first three families, the fourth family quarks will form corresponding quarkonia which will lead to spectacular signature at future lepton and hadron colliders [4]. It is interesting that within the democratic mass matrix approach, the small masses for the first three neutrinos are compatible with large mixing angles, assuming that the neutrinos are of the Dirac type [5]. There are a number of papers devoted to the indirect manifestations of the fourth family in B-meson decays (see, for example, [6] and references therein), however, there are a lot of SM extensions which lead to the similar consequences. In principle, the fourth family quarks can be observed at Tevatron (before LHC run) if anomalous $q_4 q V$ interaction exist [7].

2. SUSY vs Compositness [8]

In is known that even the three family MSSM contains the huge number of (observable!) free parameters. On the other hand, there are serious arguments (including "historical" ones) favoring the existence of a new level of compositness. Therefore, SUSY should be realized at the most fundamental pre(pre)onic level and the spectrum of "supersymmetric partners" of SM particles may be quite different comparing to MSSM expectations. Moreover, "right" sneutrino can be the lighest supersymmetric particle. Then, in the case of three SM families flavor democracy favors large value of the $\tan\beta \sim m_t/m_b$, otherwise, the fourth "MSSM" family should exist.

3. TeV Scale Lepton-Hadron and Photon-Hadron Colliders.

Today, linac-ring type ep and eA colliders (see reviews [9] and references therein) seems to be the sole realistic way to TeV scale in lepton-hadron collisions. Using the Compton backscattering of laser photons on high energy electron beam from linac [10], these machines can be converted into TeV scale $\gamma p$ colliders [11]. THERA [12] is the most advanced proposal on the subject. The idea of QCD Explorer (70 GeV "CLIC" on LHC) was arised at the informal meeting held at CERN last summer [13] in order to discuss the possibility to intersect CLIC with LHC. A comparison of e-linac and e-ring versions of the LHC and VLHC based ep colliders had been performed in [14].

4. Linac-Ring Type Factories: TAC Project

Linac-ring type $e^+e^-$ colliders can be considered as candidates for high luminosity particle factories. Today, the B-factory option [15] lose its attractiveness with coming into operation of KEK-B and PEP-B asymmetric ring-ring type machines. Moreover, a super-high luminosity, $10^{36}$ cm$^{-2}$s$^{-1}$, ring-ring type B-factories are discussed. Nevertheless, linac-ring type c-τ- [16] and φ- [17] factories are still the matter of interest. The last one is considered as the part of Turkic Accelerator Center (TAC) project (for details, see http://bilge.science.ankara.edu.tr). The c-τ-factory can be considered as the next stage of the TAC project.

Acknowledgements. We are grateful to organizers of ICFA seminar, especially to Professor L. Maiani, for invitation. This work is partially supported by the Turkish State Planning Organization (DPT) under the Grant No 2002K120250.